\documentclass[10pt,a4paper]{article}

\usepackage{alltt}
\usepackage{amsmath}
\usepackage{amssymb}
\usepackage{amsthm}
\usepackage[english]{babel}
\usepackage{booktabs}
\usepackage{cite}
\usepackage{color}
\usepackage{footnote}
\usepackage[margin=25mm,includefoot]{geometry}
\usepackage{graphicx}
\usepackage[pdftex]{hyperref}
\usepackage[utf8]{inputenc}
\usepackage{listings}
\usepackage[expansion=true,kerning=true,protrusion=true]{microtype}
\usepackage{multirow}
\usepackage{picins}
\usepackage{subfig}
\usepackage{xspace}

\definecolor{lightgray}{gray}{0.97}
\definecolor{forestgreen}{RGB}{0,128,0}
\definecolor{navyblue}{RGB}{0,0,128}
\definecolor{brick}{RGB}{128,0,0}
\definecolor{prepro}{RGB}{128,0,128}

\theoremstyle{definition}%
\newtheorem{problem}{Problem}
\newtheorem{example}{Example}
\newtheorem{solution}{Solution}

\providecommand{\symdim}{\ensuremath{d}\xspace}
\providecommand{\symdimmax}{\ensuremath{d_{\max}}\xspace}
\providecommand{\symtypeset}[1]{\ensuremath{\mathcal{T}_{#1}}}

\providecommand{\symtype}[2]{\ensuremath{t_{#1}^{#2}}}

\providecommand{\symtypenum}[1]{\ensuremath{n_{#1}}}

\providecommand{\symcondset}[1]{\ensuremath{\mathcal{C}_{#1}}\xspace}
\providecommand{\symcond}[2]{\ensuremath{c_{#1}^{#2}}}
\providecommand{\symcondx}[2]{\symcondx{#1}{#2}\xspace}
\providecommand{\symtemparg}[1]{\ensuremath{t_{#1}}}

\providecommand{\symfunctor}{\ensuremath{\mathtt{f}}}
\providecommand{\symfunctorx}{\symfunctor\xspace}
\providecommand{\symcbprod}{\ensuremath{\mathcal{N}_{\Pi}}\xspace}
\providecommand{\symcbsum}{\ensuremath{\mathcal{N}_{\Sigma}}\xspace}
\providecommand{\symiter}[2]{\ensuremath{\tau_{#1}^{#2}}}

\providecommand{\symiterend}[1]{\ensuremath{\tau_{#1}^{n_i+1}}}

\providecommand{\symidb}{\ensuremath{\mathit{id}}}
\providecommand{\symidbx}{\symidb\xspace}
\providecommand{\symid}[1]{\ensuremath{\symidb_{#1}}}

\providecommand{\range}[1]{\ensuremath{1,\ldots,#1}\xspace}
\providecommand{\rangetypenumi}{\ensuremath{\range{\symtypenum{i}}}\xspace}
\providecommand{\rangecondnumi}{\ensuremath{\range{\symtypenum{i}-1}}\xspace}

\providecommand{\forid}{\texttt{for\_id}\xspace}

\providecommand{\foreach}{\texttt{for\_each}\xspace}
\providecommand{\pos}{\texttt{pos}\xspace}

\providecommand{\casex}[1]{\ensuremath{#1}}
\providecommand{\caseA}{\casex{C_{16}^f}}
\providecommand{\caseB}{\casex{C_{32}^f}}
\providecommand{\caseC}{\casex{C_{16}^d}}
\providecommand{\caseD}{\casex{C_{32}^d}}
\providecommand{\caseE}{\casex{C_*^*}}
\providecommand{\caseEf}{\casex{C_*^f}}
\providecommand{\caseEd}{\casex{C_*^d}}

\begin{document}

\renewcommand*{\sectionautorefname}{Section}
\newcommand{\problemautorefname}{Problem}
\newcommand{\exampleautorefname}{Example}

\title{Fake Run\hspace{.04em}-\hspace{-.2em}Time Selection of Template Arguments in C++}

\author{%
Daniel Langr\footnote{E-mail: \href{mailto:langrd@fit.cvut.cz}{langrd@fit.cvut.cz}} \ and Pavel Tvrd{\' \i}k \\
\small Czech Technical University in Prague \\
\small Department of Computer Systems \\
\small Faculty of Information Technology \\
\small Th\'{a}kurova 9, 160 00, Praha, Czech Republic \\
\and
Tom{\' a}{\v s} Dytrych and Jerry P. Draayer \\
\small Louisiana State University \\
\small Department of Physics and Astronomy \\
\small Baton Rouge, LA 70803, USA
}

\date{}

\maketitle

\begin{abstract}
C++ does not support run-time resolution of template type arguments. To circumvent this restriction, we can instantiate a template for all possible combinations of type arguments at compile time and then select the proper instance at run time by evaluation of some provided conditions. However, for templates with multiple type parameters such a solution may easily result in a branching code bloat. We present a template metaprogramming algorithm called \forid that allows the user to select the proper template instance at run time with theoretical minimum sustained complexity of the branching code.

\textbf{Keywords:} C++, run-time selection, template arguments, template metaprogramming, type sequences

\end{abstract}

\section{Introduction}
\label{sec:intro}

C++ templates allow to define a parametrized piece of code for which data types are specified later as template arguments. According to the C++ Standard~\cite{RefWorks:73}, template arguments must be known at compile time. There are, however, situations where we might want to postpone the choice of template arguments to run time. Consider, e.g., the following ones:
\begin{description}

\item[Run-time choice of floating-point precision:] Many pieces of today's scientific and engineering software allow programmers to choose the floating-point precision at compile time~\cite{RefWorks:74,RefWorks:78,RefWorks:58,RefWorks:35,RefWorks:5}. If we then want to alternate single-precision and double-precision computations, we need either to recompile programs frequently or to maintain both versions simultaneously.

\item[Minimization of memory requirements:] Indexes pointing into arrays of different sizes constitute essential parts of data structures in scientific and engineering software. Let us have an array of size $\xi$ whose elements are indexed from 0 to $\xi-1$. The minimum number of bits of the unsigned integer data type that is capable to index such an array on a 64-bit computer is then

\begin{equation}
\label{eq:1}
b(\xi)=\min\bigl\{\eta\in\{8,16,32,64\}:\,\xi\leq 2^\eta\bigr\}.
\end{equation}

In some software, such as PETSc~\cite{RefWorks:78}, users can choose between 32-bit and 64-bit data types for indexes. However, the choice has to be done at compile time and the same data type is then used for all indexes independently of the actual size of indexed arrays.

\item[Reading data from binary files:] The program might not know which data types to use until it opens the file at run time. For instance, when reading files based on the HDF5 file format~\cite{RefWorks:38}, we can find out information about the types of stored data sets in the form of numerical constants.

\end{description}

\noindent
Let us now define the problem we want to address.
\begin{problem}
\label{problem:1}

Assumptions:
\begin{enumerate}
\item Suppose \symfunctorx is a function object%
\footnote{A \emph{function object}, or simply a \emph{functor}, is an object of a class that overloads the \emph{function call operator.} See, for example, Prata~\cite{RefWorks:81} or Stroustrup~\cite{RefWorks:79} for more details.}
with a templated function call operator. We will call the number of its template parameters the \emph{dimension} of the problem and denote it by \symdim.

\item Let us have \symdim finite sequences of data types $\symtypeset{1},\ldots,\symtypeset{d}$, where
\begin{displaymath}
\symtypeset{i}=\{\symtype{i}{k}:k=\rangetypenumi\}.
\end{displaymath}

\item Let us further have \symdim sequences of mutually exclusive Boolean conditions $\symcondset{1},\ldots,\symcondset{d}$, where
\begin{displaymath}
\symcondset{i}=\{\symcond{i}{k}:k=\rangecondnumi\},
\end{displaymath}
that cannot be evaluated until run time. 

\end{enumerate}
We want to \emph{apply} the function object \symfunctorx, that is, to call 
%
\begin{lstlisting}
$\symfunctor$.operator()<$\symtemparg{1}$, ... , $\symtemparg{d}$>();
\end{lstlisting}
where
\begin{displaymath}
\symtemparg{i}=\begin{cases}
\symtype{i}{k} & \text{if there exists $k\in\{\rangecondnumi\}$ such that \symcond{i}{k} is true}, \\
\symtype{i}{\symtypenum{i}} & \text{otherwise.}
\end{cases}
\end{displaymath}

\vspace{-1.2em} \ \qed
\end{problem}

\vspace{-0.3em}
A simple one-dimensional example of \autoref{problem:1} is the run-time selection of the floating point precision for some algorithm via a program's command line option.

\begin{example}
\label{example:1}
Consider a function object called \texttt{algorithm} defined as follows:%
%
{
\begin{lstlisting}
struct {
    template <typename T>
    void operator()() {
        ... // some code
    }
} algorithm;
\end{lstlisting}
}
\noindent
Let $\symtypeset{1}=\left\{\mathtt{float,double}\right\}$ and $\symcondset{1}=\{$\verb|strcmp(argv[1], "1")|$\}$.

\ \qed
\end{example}
That is, we want to invoke \texttt{algorithm.operator()<double>()} if the value of the first command line options is \texttt{"1"}, and \texttt{algorithm.operator()<float>()} otherwise. The obvious solution is to branch the code according to the provided condition as follows:
\begin{lstlisting}
if (strcmp(argv[1], "1")) 
    algorithm.operator()<float>(); 
else
    algorithm.operator()<double>();
\end{lstlisting}

Generally, for \autoref{problem:1}, we may define the solution based on code branching as follows.
\begin{solution}\ 
 
\begin{lstlisting}
if ($\symcond{1}{1}$ && $\symcond{2}{1}$ && $\cdots$ && $\symcond{d}{1}$)
    $\symfunctor$.operator()<$\symtype{1}{1}$, $\symtype{2}{1}$, ... , $\symtype{d}{1}$>();
else if ($\symcond{1}{1}$ && $\symcond{2}{1}$ && $\cdots$ && $\symcond{d}{2}$)
    $\symfunctor$.operator()<$\symtype{1}{1}$, $\symtype{2}{1}$ , ... , $\symtype{d}{2}$>();
...
else
    $\symfunctor$.operator()<$\symtype{1}{\symtypenum{1}}$, $\symtype{2}{\symtypenum{2}}$, ... , $\symtype{d}{\symtypenum{d}}$>();
\end{lstlisting}

\vspace{-.4em} \ \qed
\end{solution}

\vspace{-.6em}
The drawback of this solution is obvious---the complexity of the branching code grows combinatorially (recall that we need the templated function call operator to be instantiated for all possible combinations of data types). That is, it yields the total number of branches $\symcbprod=\prod_{i=1}^{d}\symtypenum{i}$.

The lower bound on the number of branches is $\symcbsum=\sum_{i=1}^d\symtypenum{i}$, because all conditions must be evaluated in the worst case scenario. This lower bound could be achieved by the following \emph{imaginary} code:
\begin{lstlisting}
???? $\symtemparg{1}$;
(* \vspace{-.6em} *)
if ($\symcond{1}{1}$)
    $\symtemparg{1}$ = $\symtype{1}{1}$;
else if ($\symcond{1}{2}$)
    $\symtemparg{1}$ = $\symtype{1}{2}$;
...
else if ($\symcond{1}{\symtypenum{1}-1}$)
    $\symtemparg{1}$ = $\symtype{1}{\symtypenum{1}-1}$;
else
    $\symtemparg{1}$ = $\symtype{1}{\symtypenum{1}}$;
(* \vspace{-.6em} *)
... // similarly for the remaining dimensions
(* \vspace{-.6em} *)
$\symfunctor$.operator()<$\symtemparg{1}$, $\symtemparg{2}$, ... , $\symtemparg{d}$>();
\end{lstlisting}
%
%
%
Unfortunately, such a code is not valid, because C++ does not allow to assign types%
\footnote{Using \texttt{typedef}s instead of assignment would not help here, because even though most C++ compilers do accept \texttt{typedef}s in a function body, such type definitions would not propagate from inside the code branches.}.

In this paper, we present a solution of \autoref{problem:1} that achieves the lower bound on the number of required code branches \symcbsum. Its application to \autoref{example:1} may look simply like:
\begin{lstlisting}
typedef boost::mpl::vector<float, double> fp_types;
...
int fp_id = (strcmp(argv[1], "1")) ? 0 : 1;
for_id<fp_types>(algorithm, fp_id);
\end{lstlisting}
This solution is based on template metaprogramming~\cite{RefWorks:51,RefWorks:84,RefWorks:82,RefWorks:83} and sequences of types from Boost Metaprogramming Library (MPL)~\cite{RefWorks:53}. Although we cannot choose template arguments at run time, we can choose positions (ids) of desired data types inside type sequences. Based on these ids, the presented \forid algorithm invokes the desired template instance%
\footnote{In fact, \texttt{for\_id} is an ordinary C++ function template (not a metafunction). However, its functionality matches the category that is called \emph{algorithms} in Boost MPL.}%
. Since both approaches are syntactically similar, we refer to our solution as \emph{fake} run-time selection of template arguments.

The rest of the paper is organized as follows. In \autoref{sec:relwork}, the previous work related to our problem is presented and analyzed. \autoref{sec:design} covers design and implementation of the proposed \forid algorithm. In \autoref{sec:results}, experiments are described and their results are presented and discussed. \autoref{sec:concl} summarizes the properties of the \forid algorithm and describes its usage in an existing high performance computing (HPC) code.

Note that this paper is an extended version of our previous work~\cite{RefWorks:107}. The additional material includes the following items:
\begin{itemize}

\item We tested the compatibility of the \forid algorithm with various C++ compilers. The results of these experiments are presented and discussed in \autoref{sec:results}.

\item We measured the dependence of compilation time and memory requirements of the \forid algorithm on the problem dimension and on the length of type sequences. The results of these experiments are presented and discussed in \autoref{sec:results}.

\item \autoref{sec:relwork} was extended to include additional details about the related work. 

\item Due to the space limitations, the syntax of the code excerpts is heavily compressed in~\cite{RefWorks:107}. In effect, the code presented therein requires modifications to become valid C++ code. In this paper, valid C++ code, which can be easily embedded into real programs, is shown.



\end{itemize}

\section{Related Work}
\label{sec:relwork}

C++ templates and template metaprogramming have always been intended to be utilized primarily at compile time. Boost MPL~\cite{RefWorks:53} is a widely-used general-purpose metaprogramming library advertised as \textit{``high-level C++ template meta\-programming framework of compile-time algorithms, sequences and metafunctions''}~\cite{RefWorks:85}. There are many compile-time algorithms in Boost MPL, but only one run-time algorithm: \foreach. The call 
%
\texttt{boost::mpl::for\_each<seq>(f)}
%
applies the function object \texttt{f} (calls its function call operator) to every element of the type sequence \texttt{seq} at run time. There are two significant differences between Boost MPL's \foreach and our \forid:
\begin{enumerate}

\item \foreach applies the function object to all elements in the type sequence. \forid applies the function object to a single element only---the one that is identified by its position (id).

\item \foreach is one-dimensional, that is, it can operate only on a single type sequence. \forid is multi-dimensional
and we designed it with no imposed limit of the number of dimensions (type sequences). This limit is given solely by the compiler.

\end{enumerate}

Generic Image Library (GIL)~\cite{RefWorks:86} 
allows to design generic algorithms for different types of images that are not known until run time. According to the actual image properties, such as the color space or bit depth, a proper template instance of the algorithm is invoked at run time. However, this functionality is tightly coupled with GIL and it is not presented as an independent metaprogramming algorithm for general-purpose usage. Some implementation details are described in the paper of Bourdev and J\"arvi~\cite{RefWorks:60}, but for deeper understanding of their solution, we need to study undocumented functions and class templates from GIL's source code%
\footnote{%
Looking at the \texttt{apply\_operation\_base.hpp} header file, we find that proper template instances are selected via extensive \texttt{switch} statements in the \texttt{apply} member function of the \texttt{apply\_ope\-ra\-tion\_fwd\_fn} class template. To cover numerous possibilities that may arise, this class template is multiple times partially specialized. Direct definition of these specializations would lead to an extremely large header file, thus, the Boost Preprocessor metaprogramming library is utilized here. The size of the \texttt{apply\_operation\_base.hpp} header file is only 10.5 kB, however, after removing all other included header files, we measured that the size of its preprocessed output is 2.7 MB.
}.

\section{Design and Implementation}
\label{sec:design}

\subsection{Notation}

We adhere to the following notation rules in the text below:
\begin{enumerate}

\item The header files that should be included to compile our examples are listed in \autoref{sec:hfs}.

\item We suppose that the \texttt{using} directive is provided for the \texttt{boost::mpl} namespace, that is:
\begin{lstlisting}
using namespace boost::mpl;
\end{lstlisting}


%
%
\item We use the \symiter{}{} symbol for MPL iterators such that \texttt{deref<}\symiter{i}{k}\texttt{>::type} is equal to \symtype{i}{k},
and \symiterend{i} denotes \texttt{end<}\symtypeset{i}\texttt{>}.

\end{enumerate}
By the symbol \symidbx we denote a zero-based index into a type sequence. We say that \symidbx is \emph{valid} for the sequence \texttt{S} if it belongs to $\{0,\ldots,\,$\texttt{size<S>::value}$\;-\,1\}$.

\subsection{Initial Step}
\label{sec:initstep}

Let us first define a metafunction \pos that returns a zero-based index of a type within a type sequence (that is, \pos\texttt{<}$\symtypeset{i},\symtype{i}{k}$\texttt{>} is equal to $k-1$). \\
\begin{lstlisting}
template <typename S, typename T>
struct pos : distance<
    typename begin<S>::type,
    typename find<S,T>::type
>::type { };
\end{lstlisting}
Our initial solution of the one-dimensional \autoref{problem:1} is then:
{\small
\begin{lstlisting}[numbers=left,xleftmargin=2em]
// primary template
template <                                                                        (* \label{lst:init:pts} *)
   typename S1,
   typename B1 = typename begin<S1>::type,                                        (* \label{lst:init:b} *)
   typename E1 = typename end<S1>::type                                           (* \label{lst:init:e} *) 
>
struct for_id_impl_1 { 
   template <T>                                                                   (* \label{lst:init:exs} *)
   static void execute(T& f, int id1) {
      if (pos<S1, typename deref<B1>::type>::value == id1)                        (* \label{lst:init:condo} *)
         f.template operator()<typename deref<B1>::type>();                       (* \label{lst:init:appl} *)
      else if (1 == distance<B1, E1>::value)                                      (* \label{lst:init:condt} *)
         throw std::invalid_argument("");
      else
         for_id_impl_1<S1, typename next<B1>::type, E1>::execute(f, id1);         (* \label{lst:init:rec} *)
   }                                                                              (* \label{lst:init:exe} *)
};                                                                                (* \label{lst:init:ets} *)

// partial specialization
template <typename S1, typename E1>                                               (* \label{lst:init:pss} *)
struct for_id_impl_1<S1, E1, E1> { 
   template <T> static void execute(T& f, int id1) { }
};                                                                                (* \label{lst:init:pse} *)
\end{lstlisting}
}
\noindent
It iterates over the type sequence \texttt{S1} either until the position of the actual type matches the desired \symidbx, or until the end of the sequence is reached. In the former case, the function object is applied. In the latter case, an exception is thrown. The partial specialization defined at lines~\ref{lst:init:pss}--\ref{lst:init:pse} is never reached at run time, however, it is needed to stop the recursive instantiation at compile time. 

Let us go back to our Example~\ref{example:1} where $\symtypeset{1}=\left\{\mathtt{float,double}\right\}$. What happens when we now call \texttt{for\_id\_impl\_1<}\symtypeset{1}\texttt{>::execute(algorithm, id1)} and \texttt{id1} is 1?
\begin{enumerate}

\item At lines~\ref{lst:init:b}--\ref{lst:init:e}, the default arguments are resolved, resulting in \texttt{<}$\symtypeset{1},\symiter{1}{1},\symiter{1}{3}$\texttt{>}.

\item The \texttt{execute} function, defined at lines~\ref{lst:init:exs}--\ref{lst:init:exe}, is invoked, wherein \texttt{id1} is equal to 1 and \texttt{pos<}$\symtypeset{1},\symtype{1}{1}$\texttt{>::value} is equal to 0.

\item \label{item:condmet} The condition at line~\ref{lst:init:condo} is hence not satisfied. At the same time, the condition at code line~\ref{lst:init:condt} is not satisfied either, because \verb|distance<|$\symiter{1}{1},\symiter{1}{3}$\verb|>::value| is 2. Hence, the following command is executed at line~\ref{lst:init:rec}:\\ \texttt{for\_id\_impl\_1<}$\symtypeset{1},\symiter{1}{2},\symiter{1}{3}$\texttt{>::execute(algorithm, 1)}.

\item The condition at line~\ref{lst:init:condo} is now satisfied, because both \texttt{pos<}$\symtypeset{1},\symtype{1}{2}$\texttt{>::value} and \texttt{id1} are equal to 1. Since \texttt{deref<}\symiter{1}{2}\texttt{>::type} is equal to \symtype{1}{2} which is \texttt{double}, the following command is executed at line~\ref{lst:init:appl}:\\ \texttt{algorithm.template operator()<double>()}.

\end{enumerate}
This is exactly what we wanted, i.e., to select the \texttt{<double>} instance of the function call operator of \texttt{algorithm} by a run-time parameter \texttt{id1} (zero-based index of \texttt{double} in \symtypeset{1} is 1).

What would happen if \texttt{id1} would be invalid---for example, if it would be equal to 10? Up to the point~\ref{item:condmet} in the previous list, the behavior would be the same. However, then it would run differently:
\begin{enumerate}

\item[$4'$.] The condition at code line~\ref{lst:init:condo} is not satisfied, because \texttt{id1} is equal to 10 and \texttt{pos<}$\symtypeset{1},\symtype{1}{2}$\texttt{>::value} is equal to 1. However, the condition at line~\ref{lst:init:condt} is now satisfied, since \texttt{dis\-ta\-nce<}$\symiter{1}{2},\symiter{1}{3}$\verb|>::value| is 1. We are already at the end of \symtypeset{1} and there are no more types to iterate over. Hence, the exception that indicates the wrong \texttt{id1} argument is thrown.

\end{enumerate}

\subsection{Extension to Multiple Dimensions}
\label{sec:simpleext}

The following solution for two dimensions is based on the same idea of iterating over type sequences---we just have two of them and for each one a separate \symidbx. 
{\small
\begin{lstlisting}[numbers=left,xleftmargin=2em]
// primary template
template <                                                                       (* \label{lst:md:pts} *)
   typename S1,
   typename S2,
   typename B1 = typename begin<S1>::type,
   typename B2 = typename begin<S2>::type,
   typename E1 = typename end<S1>::type,
   typename E2 = typename end<S2>::type,
   typename T1 = typename deref<B1::type>
>
struct for_id_impl_2 {
   template <typename T>
   static void execute(T& f, int id1, int id2) {
      if (pos<S1, deref<B1>::type>::value == id1)                                (* \label{lst:md:condo} *)
         for_id_impl_2< 
            S1, S2, E1, B2, E1, E2, typename deref<B1>::type
         >::execute(f, id1, id2);
      else if (1 == distance<B1, E1>::value)
         throw std::invalid_argument("");    
      else
         for_id_impl_2<
            S1, S2, typename next<B1>::type, B2, E1, E2, T1
         >::execute(f, id1, id2);
   }
};                                                                               (* \label{lst:md:pte} *)

// partial specialization #1
template <                                                                       (* \label{lst:md:pss} *)
   typename S1, typename S2, typename B2,
   typename E1, typename E2, typename T1
>
struct for_id_impl_2<S1, S2, E1, B2, E1, E2, T1> {
   template <typename T>
   static void execute(T& f, int id1, int id2) {
      if (pos<S2, typename deref<B2>::type>::value == id2)                       (* \label{lst:md:condt} *)
         f.template operator()<T1, typename deref<B2>::type>();
      else if (1 == distance<B2, E2>::value)
         throw std::invalid_argument("");
      else
         for_id_impl_2<
            S1, S2, E1, typename next<B2>::type, E1, E2, T1
         >::execute(f, id1, id2);
   }
};                                                                               (* \label{lst:md:pse} *)

// partial specialization #2
template <
   typename S1, typename S2, typename E1, typename E2, typename T1
>
struct for_id_impl_2<S1, S2, E1, E2, E1, E2, T1> {
   template <typename T>
   static void execute(T& f, int id1, int id2) { }
};
\end{lstlisting}
}
\noindent
In the primary template defined at lines~\ref{lst:md:pts}--\ref{lst:md:pte}, the program iterates over the first type sequence \texttt{S1}. When the desired type is found, that is, when the condition at line~\ref{lst:md:condo} is satisfied, the function object cannot be applied, because the second type is not yet known. The resolved type is stored into the template parameter \texttt{T1} and the process proceeds to the second dimension. This is done by setting \texttt{B1} to \texttt{E1}, which causes the transition to the first partial specialization defined at lines~\ref{lst:md:pss}--\ref{lst:md:pse}. This partial specialization iterates over the second type sequence and when \verb|id2| is matched at line~\ref{lst:md:condt}, the function object can be applied, since all data types are now known.

Extension to 3 and more dimensions can be done by following the same pattern. However, this approach has a quadratic complexity of the number of definitions. For the dimension \symdim, we need $\symdim+1$ definitions---a primary template and \symdim partial specializations. So, if we want to support all dimensions from 1 to some \symdimmax, we finally need $\symdimmax(\symdimmax+3)/2=O(\symdimmax^2)$ definitions, which is not optimal.


\subsection{The Optimal Solution}

We present here a solution that needs only $2\symdimmax+1=O(\symdimmax)$ definitions. It primarily uses only $\symdimmax+1$ definitions that are common for all $\symdim\in\{1,\ldots,\symdimmax\}$. For $\symdimmax=2$ these definitions are the following: 
{\small
\begin{lstlisting}[numbers=left,xleftmargin=2em]
// primary template
template <
   int D,
   typename S1,
   typename S2 = vector<>,
   typename B1 = typename begin<S1>::type,
   typename B2 = typename begin<S2>::type,
   typename E1 = typename end<S1>::type,
   typename E2 = typename end<S2>::type,
   typename T1 = typename deref<B1>::type,
   typename T2 = typename deref<B2>::type
>
struct for_id_impl {             
   template <typename T>
   static void execute(T& f, int id1, int id2 = 0) {
      if (pos<S1, typename deref<B1>::type>::value == id1)                               
         if (1 == D)                                                              (* \label{lst:opt:condo} *)
            executor<D, typename deref<B1>::type, T2>::execute(f);
            // f.template operator()<typename deref<B1>::type>();                 (* \label{lst:opt:commo} *)
         else
            for_id_impl<
               D, S1, S2, E1, B2, E1, E2, typename deref<B1>::type
            >::execute(f, id1, id2);
      else if (1 == distance<B1, E1>::value)                                      (* \label{lst:opt:condt} *)
         throw std::invalid_argument("");
      else
         for_id_impl<
            D, S1, S2, typename next<B1>::type, B2, E1, E2, T1
         >::execute(f,id1);
   }
};

// partial specialization #1
template <
   int D, typename S1, typename S2, typename B2,
   typename E1, typename E2, typename T1, typename T2
>
struct for_id_impl<D, S1, S2, E1, B2, E1, E2, T1, T2> {                   
   template <typename T>
   static void execute(T& f, int id1, int id2 = 0) {
      if (pos<S2, typename deref<B2>::type>::value == id2)
         executor<D, T1, typename deref<B2>::type>::execute(f);
         // f.template operator()<T1, typename deref<B2>::type>();                (* \label{lst:opt:commt} *)
      else if (1 == distance<B2, E2>::value)                                      (* \label{lst:opt:condh} *)
         throw std::invalid_argument("");
      else
         for_id_impl<
            D, S1, S2, E1, typename next<B2>::type, E1, E2, T1
         >::execute(f,id1,id2);
   }
};

// partial specialization #2
template <
   int D, typename S1, typename S2,
   typename E1, typename E2, typename T1, typename T2>                        
struct for_id_impl<D, S1, S2, E1, E2, E1, E2, T1, T2> {
   template <typename T>
   static void execute(T& f, int id1, int id2 = 0) { }
};
\end{lstlisting}
}
\noindent
The idea of iterating over type sequences and moving to the next dimension after resolving the actual one is preserved. Comparing \verb|for_id_impl| with the previously defined template \verb|for_id_impl_2|, we find the following essential differences:
\begin{enumerate}
\item The \verb|D| template parameter, equal to the number of dimensions \symdim, was introduced.

\item The \verb|S2| template parameter and the \verb|id2| function parameter have default values, because they are useless for one-dimensional problems and we do not want to force the user to specify meaningless values for them.

\item The condition at line~\ref{lst:opt:condo} was introduced, because when the type is resolved for a particular dimension, we need to select the further action according to the number of dimensions of the problem. At line~\ref{lst:opt:condo}, where the first type is already known, we need either
\begin{enumerate}
\item to apply the function object for one-dimensional problems (\texttt{D} is 1),
\item or to move to the next dimension for two-dimensional (generally more-than-one-dimensional) problems.
\end{enumerate}
As we further see, no such condition is needed for the first partial specialization, because the \verb|executor| structure is defined only for \verb|D| being equal to 1 or 2.

\item Unfortunately, within this new solution, we cannot apply the function object directly inside \verb|for_id_impl::execute|, as is suggested by the comments at lines~\ref{lst:opt:commo} and~\ref{lst:opt:commt}. The reason is that for a two-dimensional problem, we suppose a function object with a templated function call operator that has exactly two template parameters. However, in such a case, the call \\ \verb|f.template operator()<typename deref<B1>::type>()| \\ at line~\ref{lst:opt:commo} would trigger a compilation error, because no such one-parameter version of the function call operator exists. We have solved this problem by delegation of the application of the function object to a helper structure called \texttt{executor} that is defined as follows:
{\small
\begin{lstlisting}
template <int D, typename T1, typename T2>
struct executor;
(* \vspace{-.6em} *)
template <typename T1, typename T2>
struct executor<1, T1, T2> {
   template <typename T>
   static void execute(T& f) {
      f.template operator()<T1>();
   }
};
(* \vspace{-.6em} *)
template <typename T1, typename T2>
struct executor<2, T1, T2> {
   template <typename T>
   static void execute(T& f) {
      f.template operator()<T1, T2>();
   }
};
\end{lstlisting}
}

\end{enumerate}

As in Section~\ref{sec:simpleext}, the extension to 3 and more dimensions is straightforward. For each supported dimension, we need to define one specialization of \verb|for_id_impl| and one of \verb|executor|. Hence, we need $2\symdimmax+1$ definitions in total.

It might seem that this new solution introduces some overhead when compared with the one in Sections~\ref{sec:initstep} and~\ref{sec:simpleext}, because there is too much code branching. However, we need to realize that the conditions at lines~\ref{lst:opt:condo}, \ref{lst:opt:condt} and~\ref{lst:opt:condh} may be evaluated at compile time and an efficient compiler will not propagate the branching into the resulting machine code.

\subsection{Wrapping Up}

Although \verb|for_id_impl| already solves \autoref{problem:1}, we can make things more comfortable by introducing the following wrappers:
{\small
\begin{lstlisting}
// one-dimensional case
template <typename S1, typename T>
void for_id(T& f, int id1) {
   for_id_impl<1, S1>::execute(f, id1);
}
(* \vspace{-.6em} *)
// two-dimensional case
template <typename S1, typename S2, typename T>
void for_id(T& f, int id1, int id2) {
   for_id_impl<2, S1, S2>::execute(f, id1, id2);
}
\end{lstlisting}
}
\noindent
which allows to write simply \verb|for_id<seq>(f, id)| instead of\\ \verb|for_id_impl<1, seq>::execute(f, id)|.

%
%

\subsection{Summary}

With \forid, we may write the solution of \autoref{problem:1} as follows.

\begin{solution}\ 
\label{solution:2}

\begin{lstlisting}
int $\symid{1}$;
(* \vspace{-.6em} *)
if ($\symcond{1}{1}$)
    $\symid{1}$ = 0;
else if ($\symcond{1}{2}$)
    $\symid{1}$ = 1;
...
else if ($\symcond{1}{\symtypenum{1}-1}$)
    $\symid{1}$ = $\symtypenum{1}$ - 2;
else
    $\symid{1}$ = $\symtypenum{1}$ - 1;
(* \vspace{-.6em} *)
... // similarly for the remaining dimensions
(* \vspace{-.6em} *)
for_id<$\symtypeset{1}$, $\symtypeset{2}$, ... , $\symtypeset{d}$>(f, $\symid{1}$, $\symid{2}$, ... , $\symid{d}$);
\end{lstlisting}

\vspace{-.4em} \ \qed
\end{solution}
\vspace{-.6em}
Hence, this solution achieves the minimal number of code branches \symcbsum. 

\section{Experimental Results}
\label{sec:results}

\subsection{Test Program}

To evaluate \forid, we have developed a program for computing the dominant eigenvalue of a real symmetric matrix that is obtained from a file based on the Matrix Market file format~\cite{RefWorks:8}. 
The file name is specified as a program's command line option, therefore, the number of matrix rows (columns) and the number of nonzero elements are not known until run time. Within the program, the matrix is stored in the memory in the \emph{coordinate storage sparse format} using the following data structure:
{\small
\begin{lstlisting}
struct Matrix {
   uint64_t n, z;
   void *i, *j, *a;
} m;
\end{lstlisting}
}
\noindent
where
\begin{itemize}

\item \texttt{n} is the number of matrix rows;

\item \texttt{z} is the number of matrix nonzero elements;

\item \texttt{i}, \texttt{j} and \texttt{a} are arrays containing \emph{row indexes}, \emph{column indexes}, and \emph{values} of matrix nonzero elements, respectively.

\end{itemize}

\noindent
The program looks like:
{\small
\begin{lstlisting}[numbers=left,xleftmargin=2em]
(*\color{prepro}\verb|#if defined CASE_FOR_ID|*)
typedef vector<float, double> fp_types;                                (* \label{lst:test:fpt} *)
typedef vector<uint8_t, uint16_t, uint32_t, uint64_t> ind_types;       (* \label{lst:test:indt} *)
(*\color{prepro}\verb|#endif|*)

int main(int argc, char* argv[]) {
   std::ifstream ifs(argv[1]);                                         (* \label{lst:test:fop} *)
   while ('%' == ifs.peek())                                           (* \label{lst:test:headb} *)
      ifs.ignore(1024, '\n');                                          (* \label{lst:test:heade} *)
   ifs >> m.n >> m.n >> m.z;                                           (* \label{lst:test:dims} *)
   uint64_t q = boost::lexical_cast<uint64_t>(argv[2]);                (* \label{lst:test:iters} *)

   MatrixReader mr(m, ifs);                                            (* \label{lst:test:mr} *)

(*\color{prepro}\verb|#if defined CASE_F_16|*)                         (* \label{lst:test:mrab} *)
   mr.operator()<float, uint16_t>(); 
(*\color{prepro}\verb|#elif defined CASE_F_32|*)
   mr.operator()<float, uint32_t>(); 
(*\color{prepro}\verb|#elif defined CASE_D_16|*)
   mr.operator()<double, uint16_t>(); 
(*\color{prepro}\verb|#elif defined CASE_D_32|*)
   mr.operator()<double, uint32_t>();                                  (* \label{lst:test:mram} *)
(*\color{prepro}\verb|#elif defined CASE_FOR_ID|*)
   int fp_id = (strcmp(argv[3], "1")) ? 0 : 1;                         (* \label{lst:test:fpid} *)

   int ind_id = 3;                                                     (* \label{lst:test:indidb} *)
   if (m.n <= (1UL <<  8))
      ind_id = 0;
   else if (m.n <= (1UL << 16))
      ind_id = 1;
   else if (m.n <= (1UL << 32))
      ind_id = 2;                                                     (* \label{lst:test:indide} *)

   for_id<fp_types, ind_types>(mr, fp_id, ind_id);                    (* \label{lst:test:mrafi} *)
(*\color{prepro}\verb|#else|*)
(*\color{prepro}\verb|   #error No program case selected!|*)
(*\color{prepro}\verb|#endif|*)                                       (* \label{lst:test:mrae} *)

double lambda;                                                        (* \label{lst:test:lam} *)
PowerMethod pm(m, q, lambda);                                         (* \label{lst:test:pm} *)
(*\label{lst:test:tone}*)
(*\color{prepro}\verb|#if defined CASE_F_16|*)                        (* \label{lst:test:pmab} *)
   pm.operator()<float, uint16_t>(); 
(*\color{prepro}\verb|#elif defined CASE_F_32|*)
   pm.operator()<float, uint32_t>(); 
(*\color{prepro}\verb|#elif defined CASE_D_16|*)
   pm.operator()<double, uint16_t>(); 
(*\color{prepro}\verb|#elif defined CASE_D_32|*)
   pm.operator()<double, uint32_t>();                                 (* \label{lst:test:pmam} *)
(*\color{prepro}\verb|#elif defined CASE_FOR_ID|*)
   for_id<fp_types, ind_types>(pm, fp_id, ind_id);                    (* \label{lst:test:pmafi} *)
(*\color{prepro}\verb|#endif|*)                                       (* \label{lst:test:pmae} *)
(*\label{lst:test:tthree}*)
  std::cout << "Lambda: " << lambda << "\n";                          (* \label{lst:test:print} *)

  return 0;
}
\end{lstlisting}
}
\noindent
It consists of the following steps:

\begin{enumerate}

\item The file input stream \texttt{ifs} for the matrix file is opened at line~\ref{lst:test:fop}.

\item The header and comments are skipped at lines~\ref{lst:test:headb}--\ref{lst:test:heade}.

\item The number of rows/columns and the number of nonzero elements are read at line~\ref{lst:test:dims}.

\item The number of iterations for the power method is obtained from the second command line option at line~\ref{lst:test:iters}.

\item At line~\ref{lst:test:mr}, the \texttt{mr} function object is defined. It is responsible for allocating the arrays \texttt{m.i}, \texttt{m.j}, \texttt{m.a} and for filling their values.

\item The \texttt{mr} function object is applied at lines~\ref{lst:test:mrab}--\ref{lst:test:mrae}. Details are described further in the text.

\item The variable \texttt{lambda} for storing the resulting eigenvalue is defined at line~\ref{lst:test:lam}.

\item At line~\ref{lst:test:pm}, the \texttt{pm} function object is defined. It is responsible for computing the eigenvalue and deallocating the arrays.

\item The \texttt{pm} function object is applied at lines~\ref{lst:test:pmab}--\ref{lst:test:pmae}. Details are described further in the text.

\item The computed eigenvalue is printed out at line~\ref{lst:test:print}.

\end{enumerate}

Since we wanted to evaluate the \texttt{for\_id} algorithm, we created multiple instances (cases) of the program that are listed in \autoref{tab:cases}. In the cases \caseA--\caseD, the function call operators are invoked directly at lines~\ref{lst:test:mrab}--\ref{lst:test:mram} and~\ref{lst:test:pmab}--\ref{lst:test:pmam}, which corresponds to the classical approach where data types are resolved at compile time. However, in the case \caseE, the \forid algorithm was utilized (lines~\ref{lst:test:mrafi} and~\ref{lst:test:pmafi}), where:
\renewcommand{\arraystretch}{1.15}
\renewcommand{\tabcolsep}{2.6mm}
\begin{table}[t]
\caption{Cases of the test program.}
\begin{center}
\begin{tabular}{ccc}
\toprule
        & \multicolumn{2}{c}{Data type}          \\ \cmidrule(l){2-3}
Case    & floating point    & indexing           \\ \cmidrule(r){1-1} \cmidrule(l){2-3}
\caseA & \texttt{float}     & \texttt{uint16\_t} \\
\caseB & \texttt{float}     & \texttt{uint32\_t} \\
\caseC & \texttt{double}    & \texttt{uint16\_t} \\
\caseD & \texttt{double}    & \texttt{uint32\_t} \\
\caseE & resolved by \forid & resolved by \forid \\
\bottomrule
\end{tabular}
\end{center}
\label{tab:cases}
\end{table}
\begin{enumerate}

\item the \texttt{fp\_types} and \texttt{ind\_types} type sequences are defined at lines~\ref{lst:test:fpt} and~\ref{lst:test:indt},

\item the floating-point type \symidbx is selected by the third command line option at line~\ref{lst:test:fpid},

\item the indexing type \symidbx is selected according to the number of matrix rows/columns at lines%
\footnote{The row and column indexes are integer numbers between 0 and $\mathtt{m.n}-1$, thus, we need an unsigned integer data type of width $b(\texttt{m.n})$ bits~\eqref{eq:1}.}~\ref{lst:test:indidb}--\ref{lst:test:indide}.

\end{enumerate}
\noindent
Moreover, we distinguish two sub-cases of \caseE---\caseEf\ and \caseEd\ for  \emph{single} and \emph{double} precision computation selected at run time, respectively.

Finally, we used the following definitions of the \texttt{MatrixReader} and \texttt{Power\-Method} classes:
{\small
\begin{lstlisting}[numbers=left,xleftmargin=2em]
class MatrixReader {
   public:
      MatrixReader(Matrix& m, std::ifstream& ifs) : m_(m), ifs_(ifs) { } (*\label{lst:fo:mrcons}*)

      template <typename F, typename I>
      void operator()() {
         I* i = new I[m_.z];
         I* j = new I[m_.z];
         F* a = new F[m_.z];

         for (uint64_t k = 0; k < m_.z; ++k) {
            ifs_ >> i[k] >> j[k] >> a[k];
            i[k]--; // 1- to 0-based indexing shift
            j[k]--;
         }

         m_.i = static_cast<void*>(i); (*\label{lst:fo:voidb}*)
         m_.j = static_cast<void*>(j);
         m_.a = static_cast<void*>(a); (*\label{lst:fo:voide}*)
      }

   private:
      Matrix& m_;
      std::ifstream& ifs_;
};

class PowerMethod {
   public:
      PowerMethod(const Matrix& m, uint64_t q, double& lambda) (*\label{lst:fo:pmcons}*)
         : m_(m), q_(q), lambda_(lambda) { }

      template <typename F, typename I>
      void operator()() { (*\label{lst:fo:ttwo}*)
         I* i = static_cast<I*>(m_.i);
         I* j = static_cast<I*>(m_.j);
         F* a = static_cast<F*>(m_.a);

         std::vector<F> x(m_.n, 1.0), y(m_.n);
         F lambda = 0.0;

         // power method iterations:
         do { (*\label{lst:fo:pmb}*)
            std::fill(y.begin(), y.end(), 0.0);
            for (uint64_t k = 0; k < m_.z; ++k) {
               y[i[k]] += a[k] * x[j[k]];
               if (i[k] != j[k])
                  y[j[k]] += a[k] * x[i[k]];
            }
            lambda = 0.0;
            for (uint64_t k = 0; k < m_.n; ++k)
               if (fabs(y[k]) > fabs(lambda))
                  lambda = y[k];
            for (uint64_t k = 0; k < m_.n; ++k)
               y[k] /= lambda;
            std::copy(y.begin(), y.end(), x.begin());
         } while (--q_ > 0); (*\label{lst:fo:pme}*)

         lambda_ = static_cast<double>(lambda);

         delete[] i;
         delete[] j;
         delete[] a;
   }

   private:
      const Matrix& m_;
      uint64_t q_;
      double& lambda_;
};
\end{lstlisting}}
\noindent
Note that:
\begin{enumerate}

\item We used the power method for computing the dominant eigenvalue (lines~\ref{lst:fo:pmb}--\ref{lst:fo:pme}). 

\item We used \texttt{void} pointers for storing data whose types are not known until run time (lines~\ref{lst:fo:voidb}--\ref{lst:fo:voide}).

\item We used \emph{pass-by-value} and \emph{pass-by-reference} constructor arguments to pass data \emph{to} and \emph{from} the function objects, respectively (lines~\ref{lst:fo:mrcons} and~\ref{lst:fo:pmcons}).

\end{enumerate}

\subsection{Results and Discussion}

%
We compiled the \forid-based case \caseE\ of the test program with different compilers. The results are presented in \autoref{tab:comps}. The Intel compiler required the \texttt{-std=c++0x} command line option. The Cray compiler threw multiple errors, which involved rvalue references and the \texttt{BOOST\_STATIC\_ASSERT} macro. The IBM and Microsoft compilers were not able to handle the \texttt{f.template operator()<T>()} construct and hence they were not compliant with the C++ Standard at this point%
\footnote{The workaround here might be to use a fixed-name member function instead of the function call operator. We have not tested this possibility.} (see~\cite[\S13.5/4 and \S14.2/4]{RefWorks:73} for more details).
\renewcommand{\arraystretch}{1.00}
\renewcommand{\tabcolsep}{1.6mm}
\begin{savenotes}
\begin{table}[t]
\caption{Compatibility of the \forid-based case \caseE\ of the test program with different compilers.}
\begin{center}
\begin{tabular}{cccccc}
\toprule
Compiler  & Compiler & Boost   & Processor    & Operating & Compilation \\
vendor    & version  & version & architecture & system    & errors      \\ \cmidrule(r){1-5} \cmidrule(l){6-6}
Cray      & 8.0.7    & 1.47.0  & AMD x86\_64  & Linux     & YES%
                                                                        \\
GNU       & 4.7.1    & 1.47.0  & AMD x86\_64  & Linux     & NO          \\
IBM       & 11.1     & 1.40.0  & IBM POWER7   & AIX       & YES%
                                                                        \\
Intel     & 12.1.5   & 1.47.0  & AMD x86\_64  & Linux     & NO%
                                                                        \\
Microsoft & 16.0.0.30319 01 & 1.48.0 & Intel x86\_64 & Windows & YES%
                                                                        \\
PathScale & 4.0.12.1 & 1.47.0  & AMD x86\_64  & Linux     & NO          \\
PGI       & 12.5-0   & 1.47.0  & AMD x86\_64  & Linux     & NO          \\
\bottomrule
\end{tabular}
\end{center}
\label{tab:comps}
\end{table}
\end{savenotes}

For the measurements described below, we used the GNU C++ compiler version 4.4.4. We first compared the compilation time---the results are presented in Table~\ref{tab:cts}. When the program was built completely, the compilation time of \caseE\ was 25 percent higher compared to \caseA. When the program was compiled only, the increase was 31 percent.

\renewcommand{\arraystretch}{1.00}
\renewcommand{\tabcolsep}{3mm}
\begin{table}[t]
\caption{Compilation time of the test program in seconds; average results of 10 measurements.}
\begin{center}
\begin{tabular}{r*{2}{r@{.}l}}
\toprule
Action & \multicolumn{2}{c}{\caseA} & \multicolumn{2}{c}{\caseE} \\ \cmidrule(r){1-1} \cmidrule(l){2-5} 
preprocessing, compilation, linking & 0 & 93 & 1 & 17 \\
compilation only & 0 & 76 & 0 & 94 \\
\bottomrule
\end{tabular}
\end{center}
\label{tab:cts}
\end{table}

Next, we compared the sizes of output files---the results are presented in Table~\ref{tab:fss}. The executable file size of \caseE\ is 83 percent higher compared to \caseA.

\renewcommand{\arraystretch}{1.00}
\renewcommand{\tabcolsep}{3mm}
\begin{table}[t]
\caption{Sizes of the compiled files in kilobytes.}
\begin{center}
\begin{tabular}{r*{2}{r@{.}l}}
\toprule
File & \multicolumn{2}{c}{\caseA} & \multicolumn{2}{c}{\caseE \rule[-0.5em]{0pt}{1.7em}} \\ \cmidrule(r){1-1} \cmidrule(l){2-5}
executable & 53 & 2 & 97 & 6 \rule[0em]{0pt}{1em} \\
object & 104 & 7 & 211 & 1 \\
\bottomrule
\end{tabular}
\end{center}
\label{tab:fss}
\end{table}

\begin{figure}[t]
\centering
\subfloat[]{\label{fig:mr:a}\includegraphics{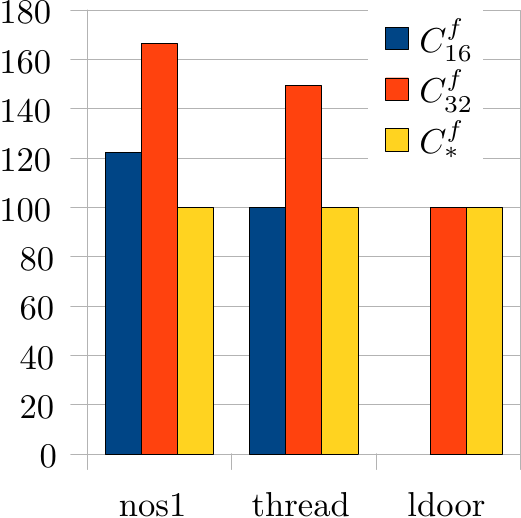}}
\hspace{4mm}
\subfloat[]{\label{fig:mr:b}\includegraphics{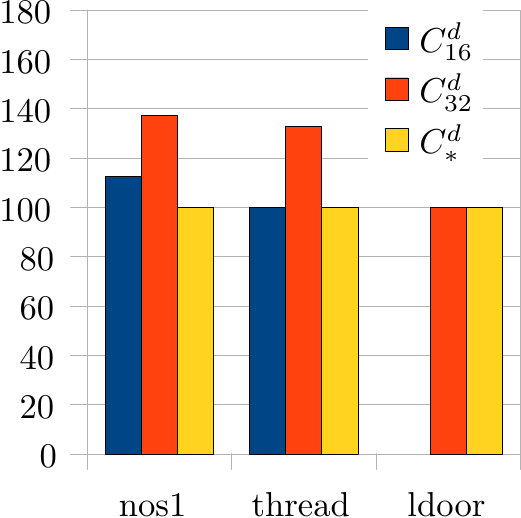}}
\caption{Comparison of program's memory requirements of the matrix and vector data structures in percents for different matrices and for computations in \emph{single}~(\protect\subref*{fig:mr:a}) and \emph{double}~(\protect\subref*{fig:mr:b}) floating-point precision. }
\label{fig:mr}
\end{figure}

For comparison of the memory requirements of the program instances, we used 3 real symmetric matrices from the University of Florida Sparse Matrix Collection~\cite{RefWorks:43}; their names and characteristics are contained in Table~\ref{tab:mats}. We measured the memory size of the matrix and vector data structures and compared them separately for single and double precision computations---the results are shown in Fig.~\ref{fig:mr}. It is clear that the program instances based on \forid always require the minimum amount of memory, because an optimal data type is used for indexes (if we included program instances using the \texttt{uint64\_t} data type into our measurements, this advantage would be even more significant).

\renewcommand{\arraystretch}{1.00}
\renewcommand{\tabcolsep}{3mm}
\begin{table}[t]
\caption{Characteristics of matrices used for experiments.}
\begin{center}
\begin{tabular}{rr *{2}{r@{.}l}}
\toprule
Number of & nos1 & \multicolumn{2}{c}{thread} & \multicolumn{2}{c}{ldoor} \\ \cmidrule(r){1-1} \cmidrule(l){2-6}
rows & 237 & 29 & $7\cdot 10^3$ & 952 & $2\cdot 10^3$ \rule[0.0em]{0pt}{1.15em} \\
nonzero elements & 627 & 2 & $3\cdot 10^6$ & 23 & $7\cdot 10^6$ \\
\bottomrule
\end{tabular}
\end{center}
\label{tab:mats}
\end{table}

Lastly, we measured the computational overhead of the \forid algorithm. We used the \texttt{clock\_gettime} POSIX function to get the actual time values in nanoseconds at 3 places:
\begin{enumerate}

\item at line~\ref{lst:test:tone} in the \texttt{main()} function,

\item at the very beginning of the function call operator of the \texttt{PowerMethod} class (line~\ref{lst:fo:ttwo})

\item at line~\ref{lst:test:tthree} in the \texttt{main()} function.

\end{enumerate}
The difference of the first and the second time values is equal to the time overhead of the \emph{invocation} of the \texttt{pm}'s function call operator. The difference of the first and the third time values is equal to the duration of the \emph{application} of the \texttt{pm}'s function call operator, that is, the whole run of the power method. The statistical information for the performed measurements are summarized in Table~\ref{tab:to}. It is clear that the time overhead introduced by the \forid algorithm is relatively high---the invocation of the function call operator takes 2.5 times longer than in the cases where this operator is called directly. However, in the context of the whole program run, this overhead is insignificant, since it is of five orders of magnitude smaller than the duration of a single power method iteration.

\renewcommand{\arraystretch}{1.00}
\renewcommand{\tabcolsep}{2mm}
\begin{table}[t]
\caption{Time differences for the \texttt{pm}'s function call operator in nanoseconds. Statistical information was gathered from 200 measurements with the thread matrix. The number of iterations of the power method was set to 10.}
\begin{center}
\begin{tabular}{cc*{3}{r@{.}l}}
\toprule
Action & Statistic & \multicolumn{2}{c}{\caseA} & \multicolumn{2}{c}{\caseD} & \multicolumn{2}{c}{\caseEf} \rule[-0.5em]{0pt}{1.7em} \\ \cmidrule(r){1-1} \cmidrule(rl){2-2} \cmidrule(l){3-8}
\multirow{3}{*}{invocation} & mean value & 191 & 1 & 191 & 0 & 477 & 9 \rule[0em]{0pt}{1em} \\
& median & 186 & 0 & 187 & 0 & 470 & 0 \\
& standard deviation & 23 & 0 & 25 & 1 & 43 & 1 \\ \cmidrule(r){1-1} \cmidrule(rl){2-2} \cmidrule(l){3-8}
\multirow{3}{*}{application} & mean value & 6 & $1\cdot 10^8$ & 6 & $1\cdot 10^8$ & 6 & $1\cdot 10^8$ \rule[0em]{0pt}{1.15em} \\
& median & 6 & $1\cdot 10^8$ & 6 & $1\cdot 10^8$ & 6 & $1\cdot 10^8$ \\
& standard deviation & 6 & $9\cdot 10^6$ & 7 & $7\cdot 10^6$ & 6 & $6\cdot 10^6$ \\
\bottomrule
\end{tabular}
\end{center}
\label{tab:to}
\end{table}


\subsection{Compilation Scalability}

Besides the experiments described heretofore, we also measured the compilation scalability of \forid. The term \emph{compilation scalability} denotes the response of a C++ compiler to the growing number of problem dimensions as well as the increasing length of type sequences. Particularly, we focused on the time and memory requirements of the compilation process. In order to carry out measurements, we created special source code:
\begin{enumerate}
\item We defined the following type sequences, each of the same length $L$:  
\begin{displaymath}
\symtypeset{i}=\{\symtype{i}{1},\ldots,\symtype{i}{L}\}\quad\text{for}\quad i=1,\ldots,d.
\end{displaymath}

\item We defined the function object \symfunctor\ as follows:
\begin{lstlisting}
struct {
    template <typename $\symtemparg{1}$, ... , typename $\symtemparg{d}$>
    void operator()() {
        ... // some code
    }
} f;
\end{lstlisting}

\item We invoked the \forid function:
\begin{lstlisting}
int $\symid{1}$, ... , $\symid{d}$; // values do not matter
for_id<$\symtypeset{1}$, ... ,$\symtypeset{d}$>(f, $\symid{1}$, ... , $\symid{d}$);
\end{lstlisting}
(Recall that we were interested in the compilation process only. Resulting programs were not executed at run time and therefore the code inside the function call operator and the values of particular type \symidb s were irrelevant here.)

\end{enumerate}

We performed the experiments for both \symdim and $L$ ranging from 1 to 6. The number of instances of the function call operator that needed to be created by the compiler within this domain is shown in \autoref{tab:instances} (it is equal to $L^d$).
\renewcommand{\arraystretch}{1.00}
\renewcommand{\tabcolsep}{3mm}
\begin{table}[t]
\caption{The number of instances of the function call operator that were created by the compiler for a range of problem dimensions and length of type sequences.} 
\begin{center}
\begin{tabular}{r*{6}{r}}
\toprule
 & \multicolumn{6}{c}{Problem dimension $d$} \\ \cmidrule(l){2-7}
Length $L$ & 1 & 2 & 3 & 4 & 5 & 6 \\ \cmidrule(r){1-1} \cmidrule(l){2-7}
1 & 1 & 1 & 1 & 1 & 1 & 1 \\
2 & 2 & 4 & 8 & 16 & 32 & 64 \\
3 & 3 & 9 & 27 & 81 & 243 & 729 \\
4 & 4 & 16 & 64 & 256 & 1024 & 4096 \\
5 & 5 & 25 & 125 & 625 & 3125 & 15625 \\
6 & 6 & 36 & 216 & 1296 & 7776 & 46656 \\
\bottomrule
\end{tabular}
\end{center}
\label{tab:instances}
\end{table}

We used the GNU C++ compiler version 4.4.6 for all experiments described in this section. To measure compilation times we utilized the \texttt{time} command available in the Linux operating system. The results of these measurements are presented in \autoref{tab:comptimes}, wherein N/A means that we were not able to obtain the corresponding value due to time/memory restrictions.
\renewcommand{\arraystretch}{1.00}
\renewcommand{\tabcolsep}{3mm}
\begin{table}[t]
\caption{Compilation times in seconds for a range of problem dimensions and length of type sequences.} 
\begin{center}
\begin{tabular}{r*{6}{r@{.}l}}
\toprule
 & \multicolumn{12}{c}{Problem dimension $d$} \\ \cmidrule(l){2-13}
Length $L$ & \multicolumn{2}{r}{1} & \multicolumn{2}{r}{2} & \multicolumn{2}{r}{3} & \multicolumn{2}{r}{4} & \multicolumn{2}{r}{5} & \multicolumn{2}{r}{6} \\ \cmidrule(r){1-1} \cmidrule(l){2-13}
1 & 0 & 4 & 0 & 4 & 0 & 4 & 0 & 4 & 0 & 4 & 0 & 4 \\
2 & 0 & 4 & 0 & 4 & 0 & 5 & 0 & 5 & 0 & 6 & 0 & 8 \\
3 & 0 & 4 & 0 & 4 & 0 & 5 & 0 & 9 & 1 & 7 & 5 & 6 \\
4 & 0 & 4 & 0 & 5 & 0 & 7 & 1 & 7 & 8 & 1 & 109 & 6 \\
5 & 0 & 4 & 0 & 5 & 1 & 0 & 3 & 9 & 57 & 3 & 2405 & 2 \\
6 & 0 & 4 & 0 & 6 & 1 & 4 & 11 & 0 & 514 & 5 & \multicolumn{2}{r}{N/A} \\
\bottomrule
\end{tabular}
\end{center}
\label{tab:comptimes}
\end{table}

One can observe that the compilation time grows rapidly with increasing values of both parameters $d$ and $L$. Moreover, the compilation time grows considerably faster than the number of instances. To show this effect, we present---in \autoref{tab:relcomptimes}---\emph{relative compilation time per instance}, which equals the ratio of the overall compilation time from \autoref{tab:comptimes} and the number of instances from \autoref{tab:instances}. There exists a compilation overhead that is not related to \forid and hence the values of relative compilation time are not relevant for small $d$ and $L$. Therefore, we restrict the results presented in \autoref{tab:relcomptimes} to those corresponding to the overall compilation time longer than 1 second.
%
%
\renewcommand{\arraystretch}{1.00}
\renewcommand{\tabcolsep}{3mm}
\begin{table}[t]
\caption{Relative compilation times per instance in milliseconds for a range of problem dimensions and length of type sequences. 
The values are presented only for overall compilation times longer than 1 second.}
\begin{center}
\begin{tabular}{r*{6}{r@{.}l}}
\toprule
 & \multicolumn{12}{c}{Problem dimension $d$} \\ \cmidrule(l){2-13}
Length $L$ & \multicolumn{2}{r}{1} & \multicolumn{2}{r}{2} & \multicolumn{2}{r}{3} & \multicolumn{2}{r}{4} & \multicolumn{2}{r}{5} & \multicolumn{2}{r}{6} \\ \cmidrule(r){1-1} \cmidrule(l){2-13}
3 & \multicolumn{2}{c}{} & \multicolumn{2}{c}{} & \multicolumn{2}{c}{} & \multicolumn{2}{c}{} & 7 & 1 & 7 & 6 \\
4 & \multicolumn{2}{c}{} & \multicolumn{2}{c}{} & \multicolumn{2}{c}{} & 6 & 5 & 7  & 9 & 26  & 8 \\
5 & \multicolumn{2}{c}{} & \multicolumn{2}{c}{} & 7 & 6 & 6 & 2 & 18 & 3 & 153 & 9 \\
6 & \multicolumn{2}{c}{} & \multicolumn{2}{c}{} & 6 & 5 & 8 & 5 & 66 & 2 & \multicolumn{2}{r}{N/A} \\
\bottomrule
\end{tabular}
\end{center}
\label{tab:relcomptimes}
\end{table}

Finally, we also measured the memory requirements of the compiler. We used the Valgrind tool---particularly its heap profiler Massif---for this experiment~\cite{RefWorks:127}. The results are presented in \autoref{tab:memory}. For $d=L=6$ the memory requirements exceeded 4~GB.
\renewcommand{\arraystretch}{1.00}
\renewcommand{\tabcolsep}{3mm}
\begin{table}[t]
\caption{Memory requirements of a compiler in megabytes for a range of problem dimensions and length of type sequences.} 
\begin{center}
\begin{tabular}{r*{6}{r@{.}l}}
\toprule
 & \multicolumn{12}{c}{Problem dimension $d$} \\ \cmidrule(l){2-13}
Length $L$ & \multicolumn{2}{r}{1} & \multicolumn{2}{r}{2} & \multicolumn{2}{r}{3} & \multicolumn{2}{r}{4} & \multicolumn{2}{r}{5} & \multicolumn{2}{r}{6} \\ \cmidrule(r){1-1} \cmidrule(l){2-13}
1 & 157 & 0 & 158 & 0 & 159 & 2 & 159 & 0 &  160 & 2 &  161 & 3 \\
2 & 158 & 0 & 160 & 3 & 162 & 3 & 166 & 7 &  173 & 1 &  185 & 0 \\
3 & 159 & 1 & 162 & 3 & 168 & 9 & 184 & 0 &  226 & 5 &  351 & 1 \\
4 & 160 & 2 & 165 & 6 & 179 & 7 & 225 & 5 &  401 & 1 &  635 & 4 \\
5 & 161 & 3 & 170 & 0 & 194 & 8 & 304 & 6 &  524 & 9 & 1842 & 8 \\
6 & 162 & 3 & 174 & 3 & 215 & 6 & 407 & 8 & 1094 & 0 & \multicolumn{2}{r}{N/A} \\
\bottomrule
\end{tabular}
\end{center}
\label{tab:memory}
\end{table}

\section{Conclusions}
\label{sec:concl}

The contribution of this paper is a new method that allows users to select data types for a piece of templated C++ code at run time with the minimal sustained complexity of code branching. The only requirement for such a piece of code is that it has to be in a form of a templated fuction call operator of some function object. 
The following conclusions can be drawn from the results of the performed experiments:
%
\begin{itemize}

\item The use of \forid allows users to select the floating-point precision for computations at run time without the need of program recompilation.

\item The use of \forid allows the best utilization of the computer memory for data structures that contain indexes. 

\item The use of \forid requires higher computational and memory resources for the compilation process, especially for longer type sequences and higher problem dimensions. 

\item The use of \forid results in a bigger executable file, that is, in a bigger program's code segment.

\item The use of \forid imposes a run-time overhead on the application of the function object.

\end{itemize}
The drawbacks seemingly prevail over the advantages. However, we need to realize that in typical real-world situations these drawbacks will be insignificant, since:
\begin{itemize}

\item Programs are usually compiled only once and then executed multiple times, and/or their compilation time is usually much smaller than their execution time.

\item The size of the code segments of running program instances are usually much smaller than the size of their data segments.

\item The execution time of the templated code is usually of several orders of magnitude longer than the run-time overhead of its invocation.

\end{itemize}

The purpose of our rather artificial test program was to evaluate the \forid algorithm.
However, we have also successfully integrated \forid into an existing HPC code, namely the code that solves \emph{symmetry-adapted no-core shell model} problems~\cite{RefWorks:23,RefWorks:24,RefWorks:25}. These problems are extremely memory-demanding and the limit for the size of the problem that can be solved on a particular HPC system is given rather by the amount of available memory than by the computational power of its processors. Inside the code, we have utilized \forid for many different tasks, including a sparse matrix-vector multiplication or a parallel file input/output of sparse matrices. 

The use of \forid allows to eliminate wasting of data memory for applications that use many different data structures containing arrays of indexes. In addition, it also allows to compile such applications only once even if the types of indexes of submitted data and/or the floating-point precision of computations vary for various runs. This may be especially useful for HPC programs that run on massively parallel supercomputers. Another example where \forid might be useful as well is the implementation of generic image algorithms as used inside GIL (see Section~\ref{sec:relwork}).


\subsection*{Acknowledgements}

This work was supported by the Czech Science Foundation under Grant No. P202/12/2011, by the U.S. National Science Foundation under Grant No. OCI-0904874, and by the U.S. Department of Energy under Grant No. DOE-0904874.

\appendix

\section{Header Files}
\label{sec:hfs}

The header files required for the \texttt{for\_id}, \texttt{for\_id\_impl}, and \texttt{execute} definitions in \autoref{sec:design}:
\begin{lstlisting}
(*\color{prepro}\verb|#include <stdexcept>|*)
(* \vspace{-.6em} *)
(*\color{prepro}\verb|#include <boost/mpl/begin_end.hpp>|*)
(*\color{prepro}\verb|#include <boost/mpl/deref.hpp>|*)
(*\color{prepro}\verb|#include <boost/mpl/distance.hpp>|*)
(*\color{prepro}\verb|#include <boost/mpl/find.hpp>|*)
(*\color{prepro}\verb|#include <boost/mpl/next_prior.hpp>|*)
(*\color{prepro}\verb|#include <boost/mpl/vector.hpp>|*)
\end{lstlisting}
We further suppose that these definitions are placed in the separate header file \texttt{for\_id.h} and that this header file is in the same directory as the test program. The header files required for the compilation of the test program defined in \autoref{sec:results} are the following:
\begin{lstlisting}
(*\color{prepro}\verb|#include <algorithm>|*)
(*\color{prepro}\verb|#include <cstring>|*)
(*\color{prepro}\verb|#include <fstream>|*)
(*\color{prepro}\verb|#include <iostream>|*)
(*\color{prepro}\verb|#include <vector>|*)
(* \vspace{-.6em} *)
(*\color{prepro}\verb|#include <stdint.h>|*)
(* \vspace{-.6em} *)
(*\color{prepro}\verb|#include <boost/lexical_cast.hpp>|*)
(* \vspace{-.6em} *)
(*\color{prepro}\verb|#include "for_id.h"|*)
\end{lstlisting}

\bibliographystyle{alphaurl}
\bibliography{langr}

\newcommand{\etalchar}[1]{$^{#1}$}
\begin{thebibliography}{DSB{\etalchar{+}}07b}

\bibitem[AG04]{RefWorks:53}
David Abrahams and Aleksey Gurtovoy.
\newblock {\em C++ Template Metaprogramming: Concepts, Tools, and Techniques
  from Boost and Beyond (C++ in Depth Series)}.
\newblock Addison-Wesley Professional, 2004.

\bibitem[Ale01]{RefWorks:51}
Andrei Alexandrescu.
\newblock {\em Modern C++ Design: Generic Programming and Design Patterns
  Applied}.
\newblock Addison-Wesley Longman Publishing Co., Inc, Boston, MA, USA, 2001.

\bibitem[BBB{\etalchar{+}}10]{RefWorks:78}
Satish Balay, Jed Brown, Kris Buschelman, Victor Eijkhout, William~D. Gropp,
  Dinesh Kaushik, Matthew~G. Knepley, Lois~Curfman McInnes, Barry~F. Smith, and
  Hong Zhang.
\newblock {PETS}c {U}sers {M}anual.
\newblock Technical Report ANL-95/11 - Revision 3.2, Argonne National
  Laboratory, 2010.

\bibitem[BJ]{RefWorks:86}
L.~Bourdev and H.~Jin.
\newblock {Generic Image Library}.
\newblock \url{http://opensource.adobe.com/gil} (accessed December, 2011).

\bibitem[BJ11]{RefWorks:60}
Lubomir Bourdev and Jaakko J\"{a}rvi.
\newblock Efficient run-time dispatching in generic programming with minimal
  code bloat.
\newblock {\em Science of Computer Programming}, 76(4):243--257, 2011.
\newblock \href {http://dx.doi.org/10.1016/j.scico.2008.06.003}
  {\path{doi:10.1016/j.scico.2008.06.003}}.

\bibitem[BN94]{RefWorks:84}
John~J. Barton and Lee~R. Nackman.
\newblock {\em Scientific and Engineering C++: An Introduction with Advanced
  Techniques and Examples}.
\newblock Addison-Wesley Longman Publishing Co., Inc., Boston, MA, USA, 1st
  edition, 1994.

\bibitem[BPR96]{RefWorks:8}
Ronald~F. Boisvert, Roldan Pozo, and Karin Remington.
\newblock The {M}atrix {M}arket {E}xchange {F}ormats: {I}nitial {D}esign.
\newblock Technical Report NISTIR 5935, National Institute of Standards and
  Technology, Dec. 1996.

\bibitem[DH11]{RefWorks:43}
T.~A. Davis and Y.~F. Hu.
\newblock The {U}niversity of {F}lorida {S}parse {M}atrix {C}ollection.
\newblock {\em ACM Transactions on Mathematical Software}, 38(1), 2011.
\newblock \href {http://dx.doi.org/10.1145/2049662.2049663}
  {\path{doi:10.1145/2049662.2049663}}.

\bibitem[DSB{\etalchar{+}}07a]{RefWorks:24}
T.~Dytrych, K.~D. Sviratcheva, C.~Bahri, J.~P. Draayer, and J.~P. Vary.
\newblock Dominant role of symplectic symmetry in ab initio no-core shell model
  results for light nuclei.
\newblock {\em Physical Review C}, 76(1):014315, 2007.
\newblock \href {http://dx.doi.org/10.1103/PhysRevC.76.014315}
  {\path{doi:10.1103/PhysRevC.76.014315}}.

\bibitem[DSB{\etalchar{+}}07b]{RefWorks:23}
T.~Dytrych, K.~D. Sviratcheva, C.~Bahri, J.~P. Draayer, and J.~P. Vary.
\newblock Evidence for symplectic symmetry in ab initio no-core shell model
  results for light nuclei.
\newblock {\em Physical Review Letters}, 98:162503, 2007.
\newblock \href {http://dx.doi.org/10.1103/PhysRevLett.98.162503}
  {\path{doi:10.1103/PhysRevLett.98.162503}}.

\bibitem[DSD{\etalchar{+}}08]{RefWorks:25}
T.~Dytrych, K.~D. Sviratcheva, J.~P. Draayer, C.~Bahri, and J.~P. Vary.
\newblock Ab initio symplectic no-core shell model.
\newblock {\em Journal of Physics {G}: Nuclear and Particle Physics},
  35(12):123101, 2008.
\newblock \href {http://dx.doi.org/10.1088/0954-3899/35/12/123101}
  {\path{doi:10.1088/0954-3899/35/12/123101}}.

\bibitem[Gen11]{RefWorks:82}
Davide~Di Gennaro.
\newblock {\em Advanced C++ Metaprogramming}.
\newblock CreateSpace, 2011.

\bibitem[GJ{\etalchar{+}}10]{RefWorks:58}
Ga\"{e}l Guennebaud, Beno\^{i}t Jacob, et~al.
\newblock Eigen, version 3.0.1, 2010.
\newblock \url{http://eigen.tuxfamily.org} (accessed July, 2011).

\bibitem[HW03]{RefWorks:35}
Michael~A. Heroux and James~M. Willenbring.
\newblock Trilinos users guide.
\newblock Technical Report SAND2003-2952, Sandia National Laboratories, 2003.

\bibitem[ISO03]{RefWorks:73}
{\em ISO/IEC 14882:2003: Programming languages: C++}.
\newblock 2003.

\bibitem[LTDD12]{RefWorks:107}
Daniel Langr, Pavel Tvrd{\'\i}k, Tom{\'a}{\v s} Dytrych, and Jerry~P. Draayer.
\newblock {Fake Run-Time Selection of Template Arguments in C++}.
\newblock In Carlo~A. Furia and Sebastian Nanz, editors, {\em Objects, Models,
  Components, Patterns (50th International Conference, TOOLS 2012)}, volume
  7304 of {\em Lecture Notes in Computer Science}, pages 140--154. Springer
  Berlin Heidelberg, 2012.
\newblock \href {http://dx.doi.org/10.1007/978-3-642-30561-0_11}
  {\path{doi:10.1007/978-3-642-30561-0_11}}.

\bibitem[NS07]{RefWorks:127}
Nicholas Nethercote and Julian Seward.
\newblock Valgrind: a framework for heavyweight dynamic binary instrumentation.
\newblock In {\em Proceedings of the 2007 ACM SIGPLAN Conference on Programming
  language Design and Implementation}, PLDI '07, pages 89--100, New York, NY,
  USA, 2007. ACM.
\newblock URL: \url{http://dx.doi.org/10.1145/1250734.1250746}, \href
  {http://dx.doi.org/10.1145/1250734.1250746}
  {\path{doi:10.1145/1250734.1250746}}.

\bibitem[Ope11]{RefWorks:74}
{\em OpenFOAM User Guide, Version 2.0.0}.
\newblock 2011.

\bibitem[Pra01]{RefWorks:81}
Stephen Prata.
\newblock {\em C++ Primer Plus (Fourth Edition)}.
\newblock Sams, Indianapolis, IN, USA, 4th edition, 2001.

\bibitem[Str00]{RefWorks:79}
Bjarne Stroustrup.
\newblock {\em {The C++ Programming Language: Special Edition}}.
\newblock Addison-Wesley Professional, 3 edition, February 2000.
\newblock URL: \url{http://www.worldcat.org/isbn/0201700735}.

\bibitem[Thea]{RefWorks:85}
{The Boost MPL Library}.
\newblock \url{http://www.boost.org/doc/libs/1_48_0/libs/mpl/doc/index.html}
  (accessed December 12, 2012).

\bibitem[Theb]{RefWorks:38}
The {HDF} {G}roup. {H}ierarchical data format version 5, 2000-2013.
\newblock \url{http://www.hdfgroup.org/HDF5/} (accessed June 3, 2013).

\bibitem[VDY05]{RefWorks:5}
Richard Vuduc, James~W. Demmel, and Katherine~A. Yelick.
\newblock {OSKI}: A library of automatically tuned sparse matrix kernels.
\newblock {\em Journal of Physics: Conference Series}, 16(1):521--530, 2005.
\newblock \href {http://dx.doi.org/10.1088/1742-6596/16/1/071}
  {\path{doi:10.1088/1742-6596/16/1/071}}.

\bibitem[VJ02]{RefWorks:83}
David Vandevoorde and Nicolai~M. Josuttis.
\newblock {\em C++ Templates---The Complete Guide}.
\newblock Addison-Wesley, 2002.

\end{thebibliography}

\clearpage

\section*{Authors' Biographies}

\parpic{\includegraphics[width=1in,clip,keepaspectratio]{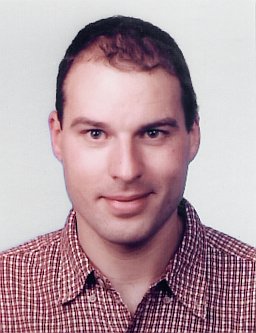}}
\noindent\textbf{Daniel Langr}
is a Ph.D. student and a researcher at the Czech Technical University in Prague, Czech Republic, where he is a member of the Parallel Computing Group at the Faculty of Information Technologies.
His main research interests include parallel I/O and visualization algorithms for sparse matrices, memory-efficient storage formats for sparse matrices, and generic programming and template metaprogramming in C++. Besides others, he focuses on discovering new approaches to high-performance-computing algorithms based on modern object-oriented technologies.
Contact him at \href{mailto:langrd@fit.cvut.cz}{langrd@fit.cvut.cz}.

\vspace{1em}
\parpic{\includegraphics[width=1in,clip,keepaspectratio]{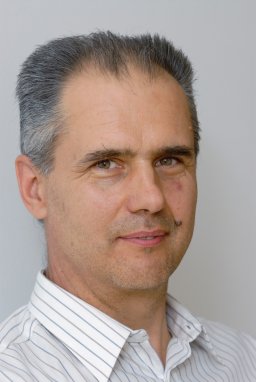}}
\noindent\textbf{Pavel Tvrd\'{i}k}
received his M.Eng. and Ph.D. in Computer Science from the Czech Technical University in Prague, in 1980 and 1991, respectively. Currently, he is a full professor at the Department of Computer Systems, Faculty of Information Technology, Czech Technical University in Prague.
His research interests include parallel computer architectures and algorithms and cluster computing.
Contact him at \href{mailto:pavel.tvrdik@fit.cvut.cz}{pavel.tvrdik@fit.cvut.cz}.

\vspace{5em}
\parpic{\includegraphics[width=1in,clip,keepaspectratio]{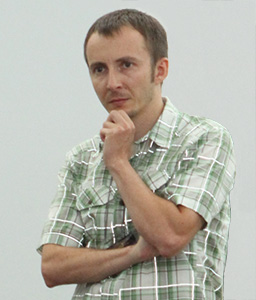}}
\noindent\textbf{Tom\'{a}\v{s} Dytrych}
is a senior research associate at the Department of Physics and Astronomy, Louisiana State University, USA.
His research interests include physics of atomic nuclei, computational group theory, general-purpose computing on graphics processing units, and application of modern programming techniques based on template capabilities of C++ for high performance computing of nuclear structure.
Contact him at \href{mailto:tdytrych@phys.lsu.edu}{tdytrych@phys.lsu.edu}.

\vspace{2.6em}
\parpic{\includegraphics[width=1in,clip,keepaspectratio]{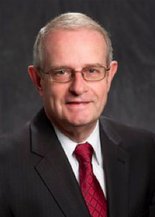}}
\noindent\textbf{Jerry P. Draayer}
is the president and CEO of the Southwestern Universities Research Association (SURA) and the Roy P. Daniels Professor of Physics at the Department of Physics and Astronomy, Louisiana State University, USA.
His research interest include the low-energy structure of atomic nuclei. Another area of interest is nonlinear phenomena, in particular, analytic and numeric tools for dealing with nonlinear processes, consequences of nonlinear couplings in physical systems, and how nonlinear effects are manifest in atomic nuclei.
Contact him at \href{mailto:draayer@phys.lsu.edu}{draayer@phys.lsu.edu}.

\end{document}